# Evolution of electronic states in n-type copper oxide superconductor via electric double layer gating

Kui Jin[1,2,3], Wei Hu[1], Beiyi Zhu[1], Dohun Kim[3,4], Jie Yuan[1], Tao Xiang[1,2], Michael S. Fuhrer[3,5], Ichiro Takeuchi[6], Richard. L. Greene[3]

[1]Beijing National Laboratory for Condensed Matter Physics, Institute of Physics, Chinese Academy of Sciences, Beijing 100190, China
[2]Collaborative Innovation Center of Quantum Matter, Beijing, 100190, China
[3]Center for Nanophysics and Advanced Materials and Department of Physics, University of Maryland, College Park, Maryland 20742, USA
[4]Department of Material Science and Engineering, Yonsei University, Seoul 120-749, Republic of Korea
[5] School of Physics, Monash University, Melbourne, Victoria 3800, Australia
[6]Department of Materials Science and Engineering, University of Maryland, College Park, Maryland 20742, USA

Email: kuijin@iphy.ac.cn

**Since the discovery of n-type copper oxide superconductors, the evolution of electron- and hole-bands and its relation to the superconductivity have been seen as a key factor in unveiling the mechanism of high-$T_c$ superconductors. So far, the occurrence of electrons and holes in n-type copper oxides has been achieved by chemical doping, pressure, and/or deoxygenation. However, the observed electronic properties are blurred by the concomitant effects such as change of lattice structure, disorder, etc. Here, we report on successful tuning the electronic band structure of n-type $Pr_{2-x}Ce_xCuO_4$ ($x$ = 0.15) ultrathin films, via the electric double layer transistor technique. Abnormal transport properties, such as multiple sign reversals of Hall resistivity in normal and mixed states, have been revealed within an electrostatic field in range of -2 V to +2 V, as well as varying the temperature and magnetic field. In the mixed state, the intrinsic anomalous Hall conductivity invokes the contribution of both electron and hole-bands as well as the energy dependent density of states near the Fermi level. The two-band model can also describe the normal state transport properties well, whereas the carrier concentrations of electrons and holes are always enhanced or depressed simultaneously in electric fields. This is in contrast to the scenario of Fermi surface reconstruction by antiferromagnetism, where an anti-correlation between electrons and holes is commonly expected. Our findings paint the picture where Coulomb repulsion**

**plays an important role in the evolution of the electronic states in n-type cuprate superconductors.**

The first n-type (electron-doped) copper oxide superconductor, $Nd_{2-x}Ce_xCuO_4$ (NCCO), was discovered in 1989[1, 2]. Soon after, comparisons between electric and thermoelectric transport measurements revealed that the optimally doped NCCO ($x$ = 0.15) actually contained both electron and hole charge carriers[3], later confirmed by the angular-dependent photoemission spectroscopy (ARPES)[4, 5] and the magnetoresistance quantum oscillation experiments[6, 7]. The coexistence of electron and hole charge carriers has been observed in other n-type copper oxide superconductors as well, such as the optimally doped $La_{2-x}Ce_xCuO_4$[8, 9] and $Pr_{2-x}Ce_xCuO_4$[10, 11]. Therefore, the competition between electron- and hole-bands becomes a common feature in n-type copper oxide superconductors[12].

A typical picture for this coexistence of electron- and hole-bands comes from a Fermi surface reconstruction. Here, a commensurate (π,π) spin-density-wave (SDW) or antiferromagnetism (AFM) order results in band folding of a large full hole Fermi surface, leading to a new Fermi surface with both electron- and hole-pockets at (π,0) and (π/2,π/2), respectively[13, 14]. This picture can be used to explain some intriguing properties, e.g. the kink in Hall coefficient[10], the strange metal behavior[15], the anomalous temperature dependent superfluid density[16], and the in-plane anisotropic magnetoresistance[17, 18], as a function of Ce doping. Meanwhile, AFM order and spin fluctuations have indeed been observed by neutron scattering, spanning over the underdoped regime[19] and extending to the overdoped side[20] in the ($T$, $x$) phase diagram. Thus, the evolution of electronic states by Ce dopants ($x$) is naturally associated with the Fermi surface reconstruction induced by AFM, invoking the relation between AFM and superconductivity[21]. Besides Ce substitution, other methods such as deoxygenation[22], irradiation[23], and pressure[24] have also been used to explore the two-band feature. However, in these cases the intrinsic electronic properties are always obscured by the concomitant lattice change and disorder.

Recently, the electric double-layer transistor (EDLT), technique has been developed to generate very large electric field and accumulating high-density charge carriers to $10^{14} \sim 10^{15}$ cm$^{-2}$ at the interface between a sample and an ionic liquid[25]. The EDLTs can tune electronic band structures of various materials. Namely, a gate voltage of several volts can realize the same effect as application of hundreds of volts in traditional solid gated

MOSFETs, which is prohibitive in reality due to the gate leakage current in solid-state dielectrics. A series of remarkable experiments have been carried out on materials such as $La_{2-x}Sr_xCuO_4$[26], $YBa_2Cu_3O_{7-\delta}$[27, 28], $ZrNCl$[29], $SrTiO_3$[30], and $KTaO_3$[31], where electrostatic modulation of carriers was used to demonstrate turning superconductivity on/off. Moreover, electrostatic gating using EDLT has been effective in probing emergent electronic phases and mapping comprehensive electronic phase diagrams[26, 27].

In this paper, we exploit EDLT to uncover the interrelationship between electron- and hole-bands in an *n*-type copper oxide superconductor. We study $Pr_{2-x}Ce_xCuO_4$ ($x = 0.15$) ultrathin films under electrostatically doping via ionic liquid. In the mixed state, an abnormal temperature- and field-dependent Hall resistivity is observed, which is remarkably suppressed in a positive electrostatic field. Meanwhile, the pristine positive Hall signal is reversed in the normal state. The rich Hall information in the mixed state invokes a consideration of energy dispersion near the Fermi surface, as well as the two-band feature. However, in the normal state, though the two-band model can account for both longitudinal resistivity ($\rho_{xx}$) and Hall resistivity ($\rho_{xy}$), the charge carrier concentrations of electrons and holes are always simultaneously increased/decreased in positive/negative electrostatic field, in contrast to the commonly expected anti-correlation between them, where the Fermi surface reconstruction results in reduced holes but enhanced electrons. Our results suggest that a dramatic change of electronic states arises within the regime of AFM or SDW. We therefore propose that the Coulomb repulsion must play an important role in the evolution of electronic states in this system.

The $Pr_{2-x}Ce_xCuO_4$ (PCCO, $x = 0.15$) ultrathin films were fabricated on (00*l*)-oriented $SrTiO_3$ substrates by a pulsed laser deposition technique[10]. The optimized PCCO sample (~ 7 unit cells, see Supplementary Note for discussion on effective thickness) starts to show superconductivity at 20 K, comparable to the bulk $T_c$. The transition width is broader than thicker films as is usually the case[26, 27]. After patterning into a standard Hall bar for $\rho_{xx}$ and $\rho_{xy}$ measurements as illustrated in Figure 1a, we chose DEME-TFSI as the ionic-liquid dielectric (IL) to guarantee the best tunability. The gate voltage $V_g$ is applied above the melting point of the IL (180 K for DEME-TFSI). The threshold voltage for electrochemical reaction (beyond which the resistance changed irreversibly) was found to be ~2.5 V for PCCO/DEME-TFSI; for the data reported below $|V_g| \leq 2V$ and the resistivity was reproducible on cooling and re-warming to >180 K. The electric transport data were then taken by sweeping the magnetic field at a fixed temperature $T$, so that the Hall signal could be precisely determined[32]. The measurements of $\rho_{xx}$ and $\rho_{xy}$ were carried out at designated temperatures, 20, 15, and 10 K, corresponding to onset, zero resistance, and below $T_c$, respectively.

Figure 1b shows temperature dependent resistivity $\rho_{xx}(T)$ at zero magnetic field with $V_g$ equal to -2, 0, +2 V. The resistivity is apparently tuned with the $T_c$ almost unaffected as shown in the lower inset. Compared to the value at 0 V, the resistivity is enhanced for $V_g$ = -2 V and reduced for $V_g$ =+2 V. Once the measurement in electrostatic fields was done, the resistivity was rechecked by sweeping the electric field above the melting point of DEME-TFSI. As demonstrated in the upper inset of Figure 1b, the $\rho_{xx}$ curve at 210 K after the gating experiments is reversible against the applied bias voltage, and overlap with the pristine data, which excludes an electrochemical reaction as the source of the change of electronic states in the present work.

Although the $T_c$ remains the same, there is a dramatic change in $\rho_{xy}$ at low temperatures, as tuned by the electrostatic field. In Fig. 2, $\rho_{xx}$ and $\rho_{xy}$ are displayed as $T$ (10, 15, and 20 K), $B$ (-9 to 9 T), and $V_g$ (-2, 0, +2 V). For clarity, we summarize the distinct feature of $\rho_{xx}$ and $\rho_{xy}$ as follows: i) After the superconductivity is suppressed by the magnetic field, $\rho_{xx}(B)$ is reduced from -2 to +2 V, not surprising as extrapolated from the high temperature resistivity; ii) above the upper critical field ($B_{c2}$), the normal state $\rho_{xy}$ is positive without gating, which is slightly enhanced in -2 V. It is tuned to negative in +2 V at 10 K and 15 K; iii) below $B_{c2}$, a huge peak in $\rho_{xy}$ is observed at 0 V once the sample enters into the mixed state. Moreover, the sign of the peak is reversed, i.e. positive at 15 K but negative at 10 K; iv) remarkably, the peak is retained in -2 V, but smeared out in +2 V.

The Hall resistivity manifests its complexity as 1) in the normal state, the sign reversal with decreasing temperature and in positive electric field, and 2) in the mixed state, the appearance of a huge peak and its sign reversal versus temperature in zero and negative electric fields, but suppressing in positive electric field. Since it has been argued that Hall conductivity is a better quantity to exhibit vortex dynamics in the mixed state[33], we also plot the corresponding Hall conductivity $\sigma_{xy}$ ($\equiv \frac{\rho_{xy}}{\rho_{xx}^2+\rho_{xy}^2}$) as seen in Figures 3a, 3b, and 3c. Consequently, some subtle features becomes more discernible. For instance, the tiny peak at ~0.06 T in 0, -2V (Figure 2f) becomes prominent in the $\sigma_{xy}(B)$ plot (Figure 3c), where the $\sigma_{xy}$ drops quickly and changes sign as the mixed state is entered.

In the mixed state, the Hall conductivity consists of two parts, $\sigma_{xy}^f$ from the vortex motion and $\sigma_{xy}^n$ from the quasiparticles inside the core of the vortices. According to the time-dependent Ginzburg-Landau (TDGL) equation derived from the BCS theory, there is a simple relation of $\sigma_{xy}^f \sim \gamma_2 (B_{c2} - B)/B$ near $B_{c2}$[34]. Here, $\gamma_2 \propto \frac{\partial N(\mu)}{\partial \mu}|_{\mu=E_F}$, with $N(\mu)$ the chemical potential dependent density of states and $E_F$ the Fermi energy[35]. Without loss of generality, the Hall conductivity can thus be written as $\sigma_{xy} = \sigma_{xy}^n + \frac{C_1}{B} + constant$. This expression does work for our experimental $\sigma_{xy}(B)$ as demonstrated in Figures 3a, 3b, and 3c, where the blue dashed lines are the fits (see Supplementary Information). The temperature dependent $C_1$ is plotted in Fig 3d, which has opposite sign at 10 K and 15 K in all electrostatic fields. And, the amplitude is notably suppressed at +2 V. Since the $C_1$ is associated with the energy derivative of the density states at the Fermi level, the abrupt change in both amplitude and sign in such a narrow temperature range implies a novel electronic state. As found by Sr dopants in hole-doped $La_{2-x}Sr_xCuO_4$, the sign reversal of $C_1$ is ascribed to experiencing a van Hove singularity with increasing Sr doping, resulting in a remarkable change in the density of states and affecting $T_c$ prominently[36]. In our case, such a dramatic change of $C_1(T)$ more likely originates from the competition between two bands since the $T_c$ of the electrostatic doped PCCO remains the same as discussed below.

A two-band model is thereby expected to capture the main characteristics of the normal-state $\rho_{xx}$ and $\rho_{xy}$. For a simple two-carrier Drude model, the $\rho_{xx}$ and $\rho_{xy}$ are expressed as,

$$\rho_{xx}(B) = \frac{(\sigma_h+\sigma_e)+\sigma_h\sigma_e(\sigma_h R_h^2+\sigma_e R_e^2)B^2}{(\sigma_h+\sigma_e)^2+\sigma_h^2\sigma_e^2(R_h-R_e)^2 B^2} \quad \text{and} \quad \rho_{xy}(B) = \frac{\sigma_h^2 R_h - \sigma_e^2 R_e - \sigma_h^2\sigma_e^2 R_h R_e(R_h-R_e)B^2}{(\sigma_h+\sigma_e)^2+\sigma_h^2\sigma_e^2(R_h-R_e)^2 B^2} B \quad \text{where}$$

$\sigma_i = \frac{n_i q^2 \tau_i}{|m_i^*|}$ and $R_i = \frac{1}{|n_i q|}$. The $n$, $\tau$, and $m^*$ represent charge carrier concentration, relaxation time, and effective mass, respectively. The subscript $i = h$ or $e$ correspond to the hole or electron band. The experimental data of $\rho_{xx}$ and $\rho_{xy}$ were fitted at the same time (red dashed lines in Figure 2) to arrive at the best self-consistent fitting parameters. As fitted by the two-band model, the carrier concentration falls in the range of $10^{19} \sim 10^{20}$ cm$^{-3}$, in agreement with the values estimated from the ARPES data[11]. Note that using a single band model to fit the data would give an unreasonable charge carrier concentration of ($\sim 10^{23}$ cm$^{-3}$).

We find that, $n_e$ and $n_h$ at fixed temperatures are always simultaneously enhanced in the positive electric field and reduced in the negative electric field, which is unusual in that typically an anti-correlation between $n_e$ and $n_h$ is expected from any aspect of electrostatic doping or the reconstruction of Fermi surface. Intuitively, the electric field will attract one type of charge carriers but repel the other. The evolution of a large hole Fermi surface into small hole and electron pockets, caused by the reconstruction, will lead to a competition between $n_e$ and $n_h$ as well. Thus, the simultaneous enhancement or reduction of $n_e$ and $n_h$ is unexpected. Another puzzling observation is that the electrostatic doping results in a dramatic change in Hall signal whereas the $T_c$ remains the same.

Xiang *et al*[37] have proposed that Fermi surface reconstruction by AFM or SDW is not the sole picture to account for the coexistence of electron and hole bands. In their picture, electrons and holes come from the upper Hubbard band and the Zhang-Rice singlet band (via hybridization among Cu 3*d* and O 2*p* orbits), respectively. For the parent compound (*x* = 0) the correlation energy *U*, from either the on-site Coulomb repulsion or between Cu 3*d* and O 2*p* bands, separates the electron and hole bands. *U* decreases with increasing doping, so the first appearance of the electron band (i.e., crossing the Fermi surface) will gradually pull up the hole band, resulting in the coexistence of electron and hole pockets on the Fermi surface. In this case, the tuning of *U* by electrostatic doping accounts for the unexpected correlation between $n_e$ and $n_h$. Moreover, it is pointed out that the $T_c$ is controlled by the distance (*D*) between electron and hole Fermi pockets in momentum space. Since the calculated conductivities, mean free paths, and effective masses of electrons and holes are almost equal in values as seen in Figures 4a, 4b, and 4c, a small tuning between the electron and hole bands can cause the sign reversal but keep the $T_c$ if the distance *D* is not obviously affected.

We now turn to the origin of the huge peak in $\rho_{xy}(B)$, at $V_g$ = -2 V and 0 V and *T* = 10 K and 15 K. Our reversible electrostatic tuning is not expected to introduce considerable additional pinning centers, thus, we rule out a pinning change as the primary cause[38]. The cutoff of the mean free path (*l*) by the vortex core can also possibly induce an abrupt change in $\rho_{xy}$, whereas this requires a clear difference between $l_e$ and $l_h$, which is also not the case (see Fig. 3a). Considering the vortex motion analog to that in superfluid $^4$He, Hagen et al[39] have obtained $v_p \propto (\frac{h\rho_s}{m^*} - \eta')$, where $\mathbf{v}_p$ is the flux-line velocity parallel to the applied transport current, $\rho_s$ is the superfluid density, and $\eta'$ is the viscosity factor of flux line in the normal direction. The $\mathbf{v}_p$ can generate a transverse electric field given by Faraday's law, $\mathbf{E} = -\frac{\mathbf{v}_p \times \mathbf{B}}{c}$. In this framework, the appearance of the huge peak in -2 V, 0 V could be linked to the sudden increase of the superfluid density upon entering into the mixed state, but the suppression of the peak in +2 V remains a mystery, subject to further research.

In conclusion, we have succeeded in tuning optimally doped PCCO ($x$ = 0.15) ultrathin films via EDLT and observed a complex Hall signal as a function temperature, magnetic field, as well as electric field. In the mixed state, the dramatic change of Hall conductivity invokes the presence of two-type carriers. In the normal state, although the two-band model can fit the resistivity and the Hall resistivity well at the same time, the simultaneous enhancement or reduction of electrons and holes indicates the tuning of electronic state stays within the range of the AFM or SDW phase. The $T_c$ remains constant with electrostatic doping yet a dramatic change in electronic state has been observed. To reconcile these observations, we therefore suggest that the Coulomb repulsion, tunable in the electrostatic field, is playing an important role.

**Acknowledgements**

We would like to thank A. Baker for the assistance to synthesize thin films. This research was supported by the National Key Basic Research Program of China (2015CB921000), the Strategic Priority Research Program (B) of the Chinese Academy of Sciences (XDB07020100), the National Natural Science Foundation of China Grant (11474338), the National Science Foundation (DMR-1104256 and 1410665), AFOSR-MURI (FA9550-09-1-0603), and AFOSR (FA95501410332).

**Additional information**

**Competing financial interests:** The authors declare no competing financial interests.

**Figure 1 Electric double-layer transistor (EDLT) and resistivity.** **(a)** Schematic of gate configuration for charge transfer doping with DEME-TFSI ionic liquid electrolyte. Pt wire is used as gate electrode. **(b)** Temperature dependent resistivity of optimal doping $Pr_{2-x}Ce_xCuO_4$ ($x$ = 0.15) ultrathin film. The resistivity of the normal state is enhanced in -2 V and reduced in +2 V. The $\rho$-$T$ curves are enlarged near the transition temperature $T_c$ in the lower inset. $T_c$ is almost unaffected by electrostatic field, and the onset $T_c$ = 20 K, which is comparable to the value of bulk. The transition width is broader than thicker film as usual in ultrathin film. At 210 K, the ionic liquid used is in liquid state and the gate voltage applied can be continuously changed. The upper inset shows the gate voltage dependent resistivity before (blue) and after (pink) the gating experiment. Both curves are reversible and overlap with each other, which excludes an electrochemical reaction as the source of the change of electronic states in this work.

**Figure 2 Temperature, magnetic field, and electrostatic field dependent $\rho_{xx}$ and $\rho_{xy}$ with $B \perp ab$-plane.** **(a), (c) and (e)** Magnetic field dependent resistivity in -2, 0, +2 V and at 10, 15, 20 K, respectively. The resistivity of the normal state is reduced from -2 to +2 V. **(b), (d) and (f)** Magnetic field dependent Hall resistivity in -2, 0, +2 V and at 10, 15, 20 K, respectively. In the normal state, $\rho_{xy}$ is positive without gating, slightly enhanced in -2 V but tuned to negative in +2 V at 10 K and 15 K. Once in the mixed state, a huge peak in $\rho_{xy}$ emerges at 0 V and the sign of the peak is reversed, i.e. positive at 15 K but negative at 10 K. Moreover, the peak is retained in -2 V, but smeared out in +2 V. The red dashed lines represent fittings to $\rho_{xx}$ and $\rho_{xy}$ simultaneously using the two-band model.(see text)

**Figure 3 Temperature, magnetic field, and electrostatic field dependent Hall conductivity. (a), (b) and (c)** Magnetic field dependent Hall conductivity in -2, 0, +2 V and at 10, 15, 20 K, respectively. The Hall conductivity can be fit by $\sigma_{xy} = \sigma_{xy}^n + \frac{C_1}{B} + const$ (blue dashed lines), and $C_1$ is proportional to $\frac{\partial N(\mu)}{\partial \mu}|_{\mu=E_F}$. Red dashed lines are the fittings to the normal states. In order to zoom in the feature at low magnetic field, the x axis in (a) and (b) is plotted in logarithmic scale. **(d)** Temperature and electrostatic field dependent $C_1$. Dashed lines are drawn to guide the eye. The abrupt change in both amplitude and sign in such a narrow temperature range indicates a dramatic change of electronic states.

**Figure 4 Fitting quantities and schematic diagram. (a), (b) and (c)** Relative conductivities, mean free paths, and effective masses of holes to electrons, respectively. The values of these quantities only slightly deviate from unity with varying temperature and electrostatic field. **(d)** SDW boundary (dashed line) is taken from Ref.17, where the Fermi surface reconstruction happens at $x_{FS} \sim 0.16$. The inset shows the temperature and electrostatic dependent carrier concentrations. The carrier concentrations of electrons and holes are always enhanced or depressed simultaneously in +2 V, or -2 V

respectively. The tuning of electronic state stays within the range of the AFM or SDW phase, so here we put the inset inside the SDW phase.

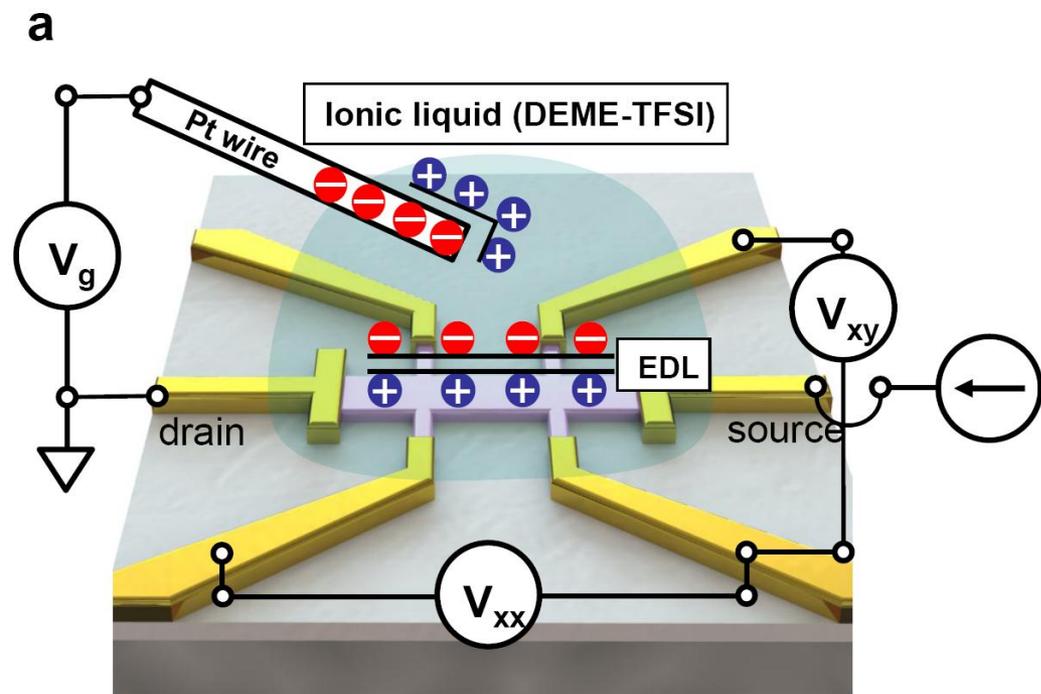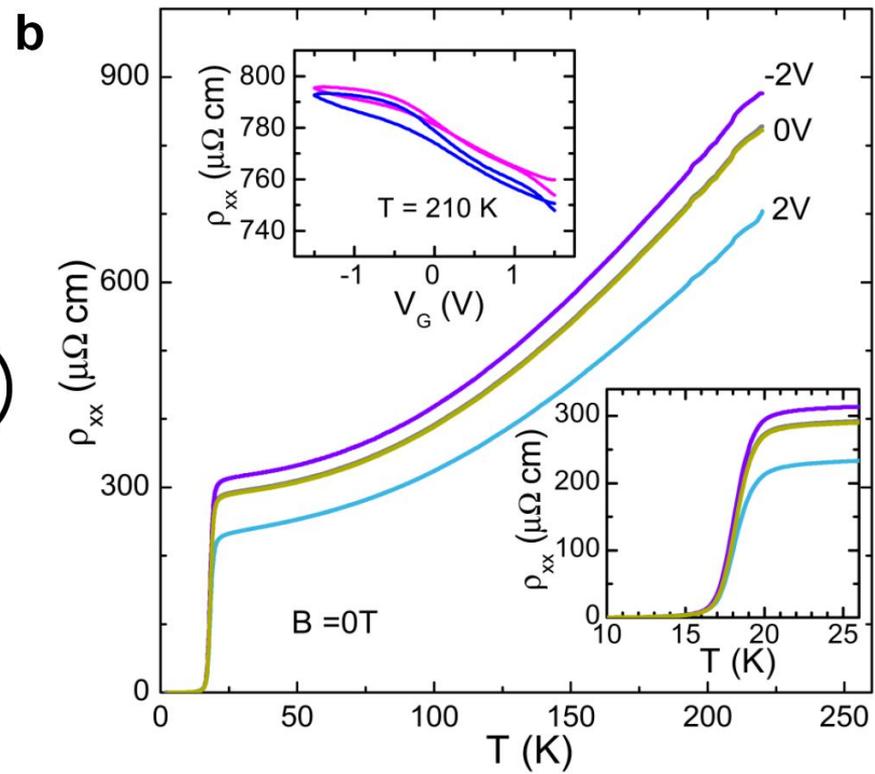

Figure 1

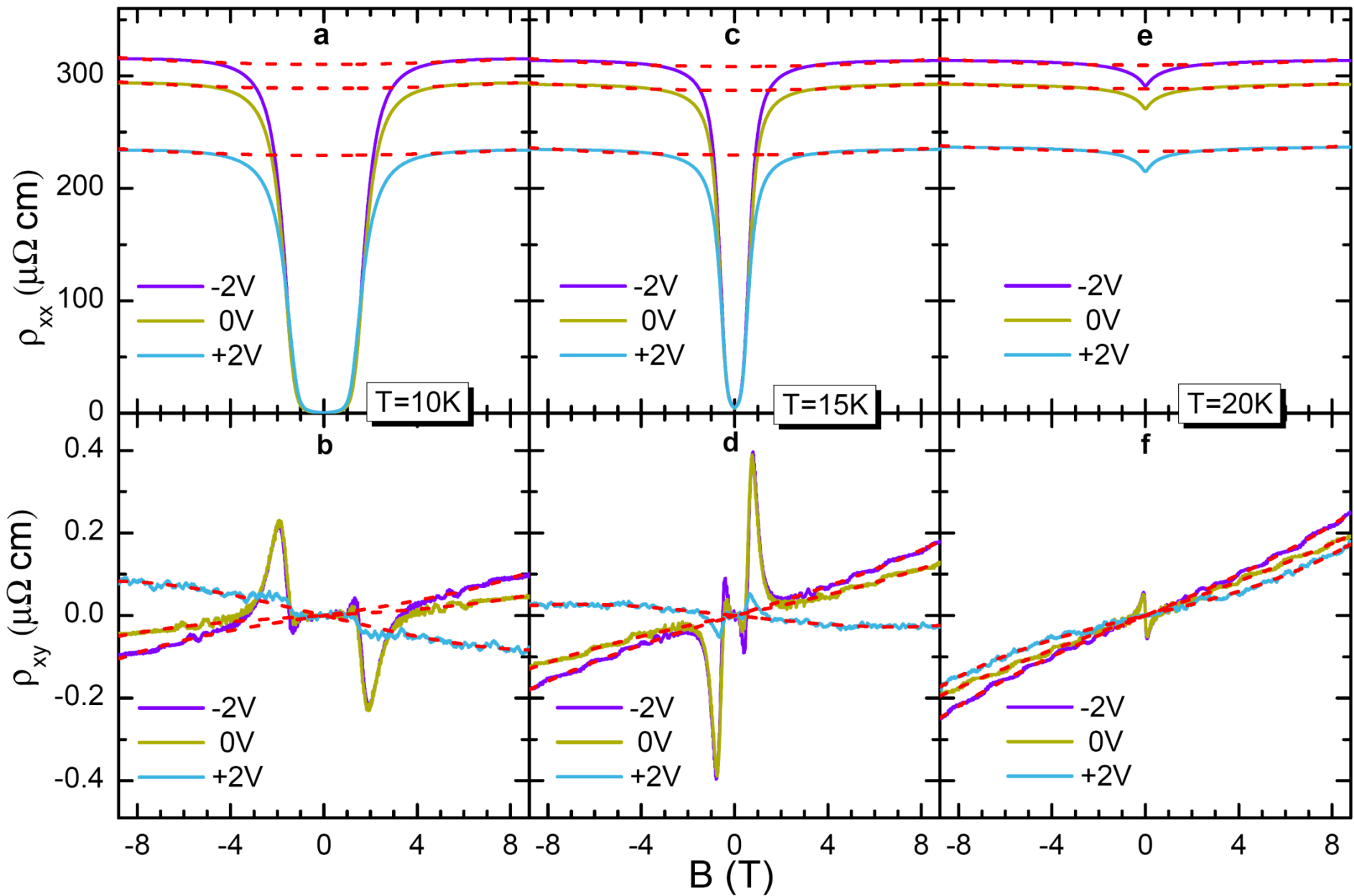

Figure 2

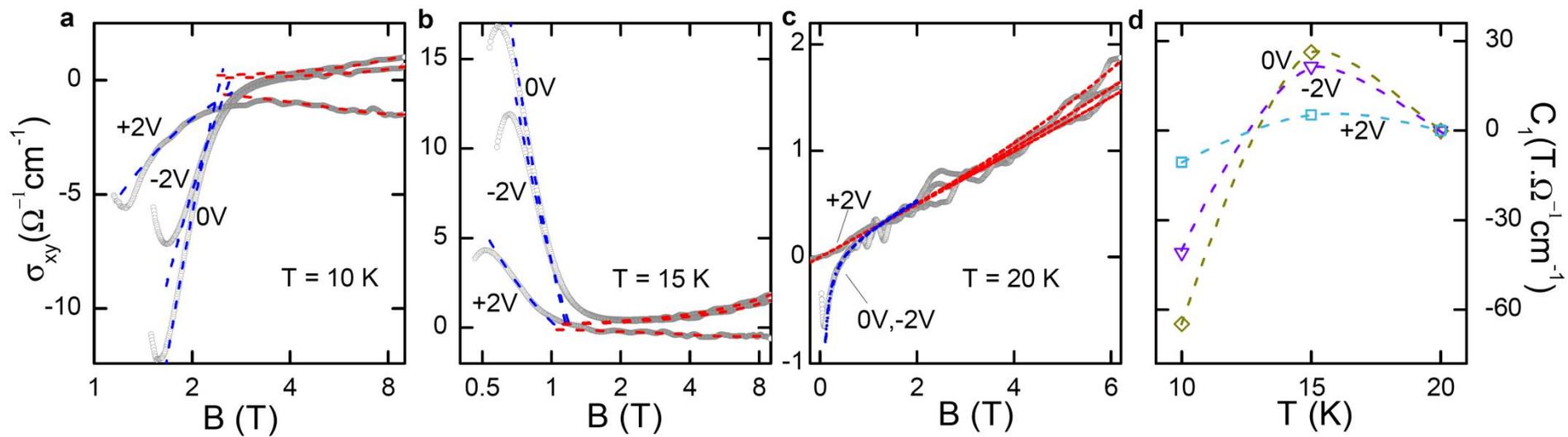

Figure 3

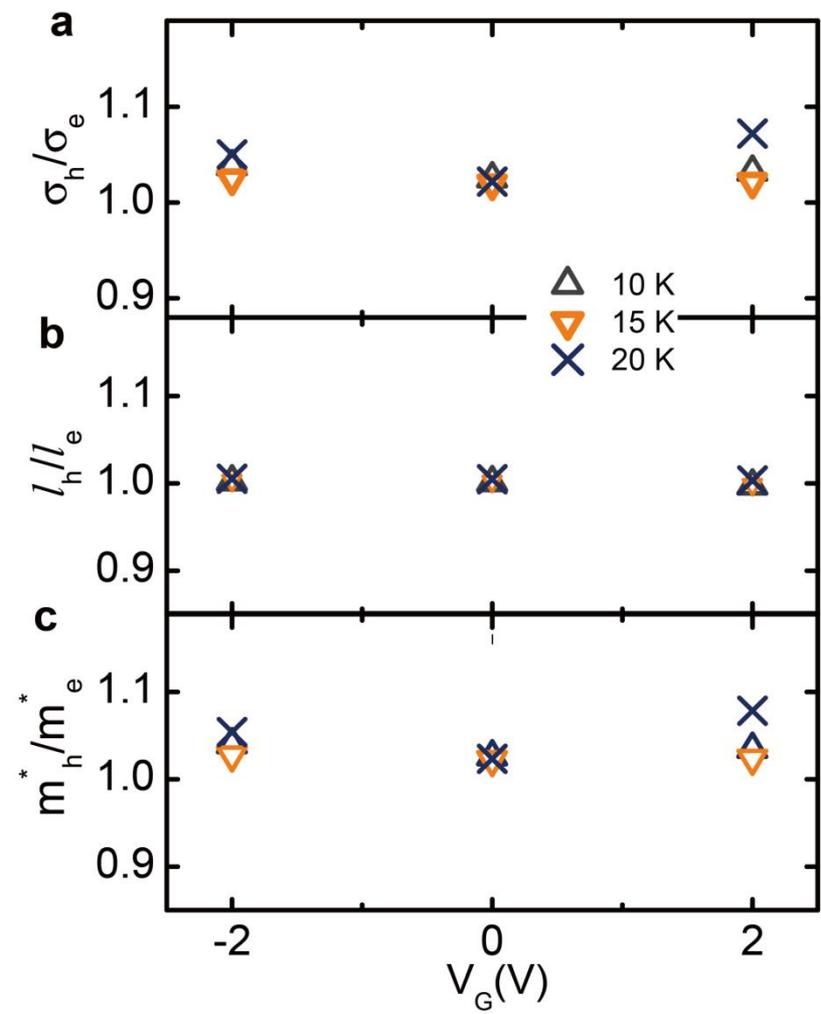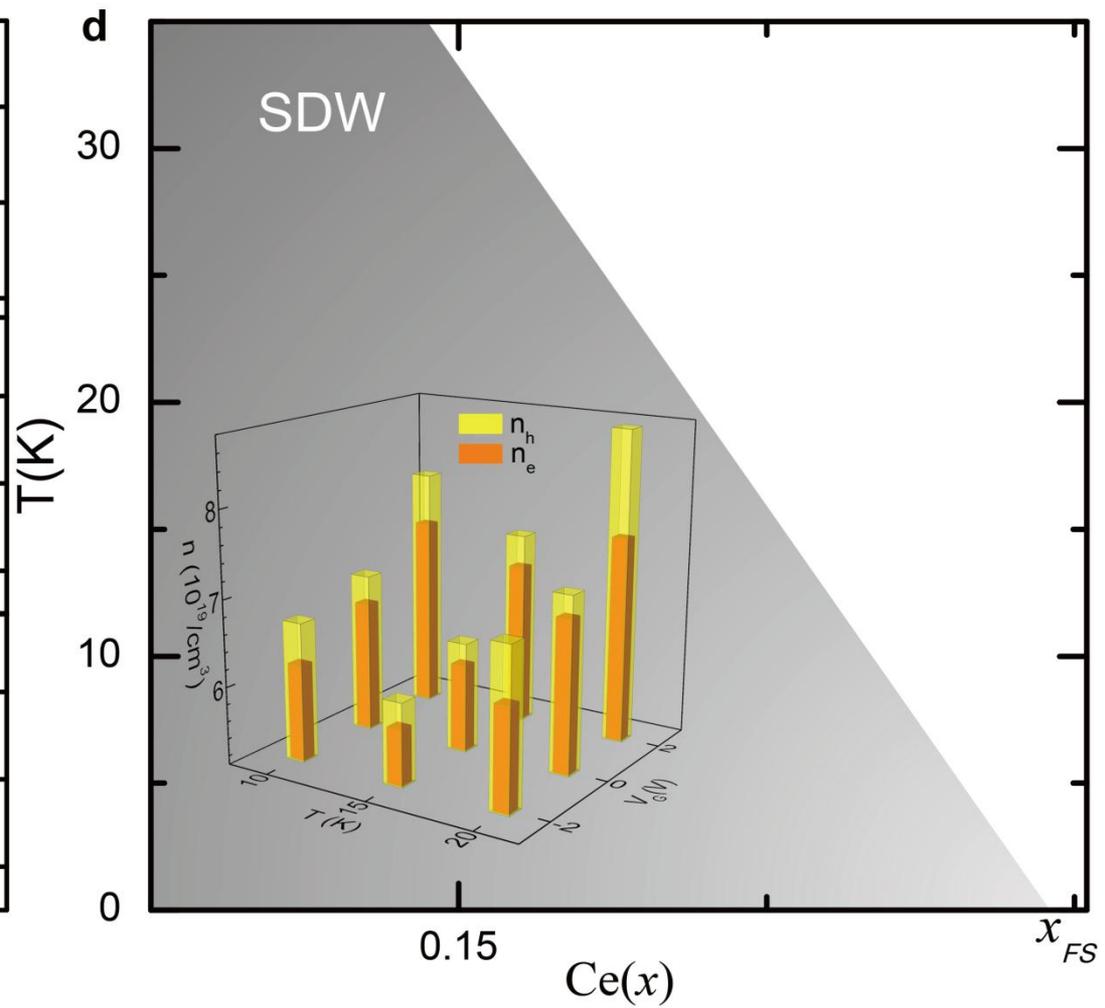

Figure 4